\documentclass[12pt]{iopart}
\usepackage{epsfig,graphicx,psfrag,amssymb,amsfonts,bbm,latexsym,color,dcolumn,bm}

\begin{document}
\title{Exploring the quantum vacuum with cylinders}

\author{F.C. Lombardo, F D Mazzitelli, and P.I. Villar}

\address{Departamento de F\'\i sica J.J. Giambiagi,
Facultad de Ciencias Exactas y Naturales, Universidad de Buenos
Aires, Ciudad Universitaria, Pabell\'on 1, 1428 Buenos Aires,
Argentina}

\ead{lombardo@df.uba.ar, fmazzi@df.uba.ar, paula@df.uba.ar}

\begin{abstract}
We review recent work on the Casimir interaction energy between
cylindrical shells. We include proposals for future experiments
involving cylinders, such us a null experiment using
quasi-concentric cylinders, a cylinder in front a conducting
plate, and a cylindrical version of the rack and pinion powered by
Casimir lateral force. We also present an exact formula for the
theoretical evaluation of the vacuum interaction energy between
eccentric cylindrical shells, and describe improved analytical and
numerical evaluations for the particular case of concentric
cylinders.
\end{abstract}

 \vspace{2pc} \noindent{\it Keywords}:
Casimir effect, quantum vacuum, cavity QED

\submitto{\JPA}

%\maketitle

%%%%%%%%%%%%%%%%%%%%%%%%%%%%%%%%%%%%%%%%%

\section{Introduction}

Up to now, most experiments aiming at a  measurement of the
Casimir force have been performed with parallel plates \cite{pp},
or with a sphere in front of a plane \cite{sp}. The parallel
plates configuration has a stronger signal, but the main
experimental difficulty is to achieve parallelism between the
plates. This problem is of course not present in the case of a
sphere in front of a plane, but its drawback is that the force is
several orders of magnitude smaller. On the theoretical side, the
evaluation of the electromagnetic force for this configuration is still an open
problem \cite{esfera}.

In this paper we will consider different configurations that
involve cylindrical surfaces. As we will see, these configurations
have both experimental and theoretical interest. In the next
Section we will describe some promising experimental proposals. In
Section 3 we will present an exact formula for the Casimir
interaction energy between two eccentric cylindrical shells, and
we will show that particular known formulas, such as the
corresponding to a cylinder in front of a plane or the one of
concentric cylinders, are included therein. 
Finally, for the particular case of concentric cylinders, in
Section 4 we will present a new analytic result beyond the
Proximity Force Approximation (PFA), and an improved numerical
method to evaluate the interaction energy at small distances.

%%%%%%%%%%%%%%%%%%%%%%%%%%%%%%%%%%%%%%%%%

\section{Experimental proposals}

\begin{figure}
\centerline{\psfig{figure=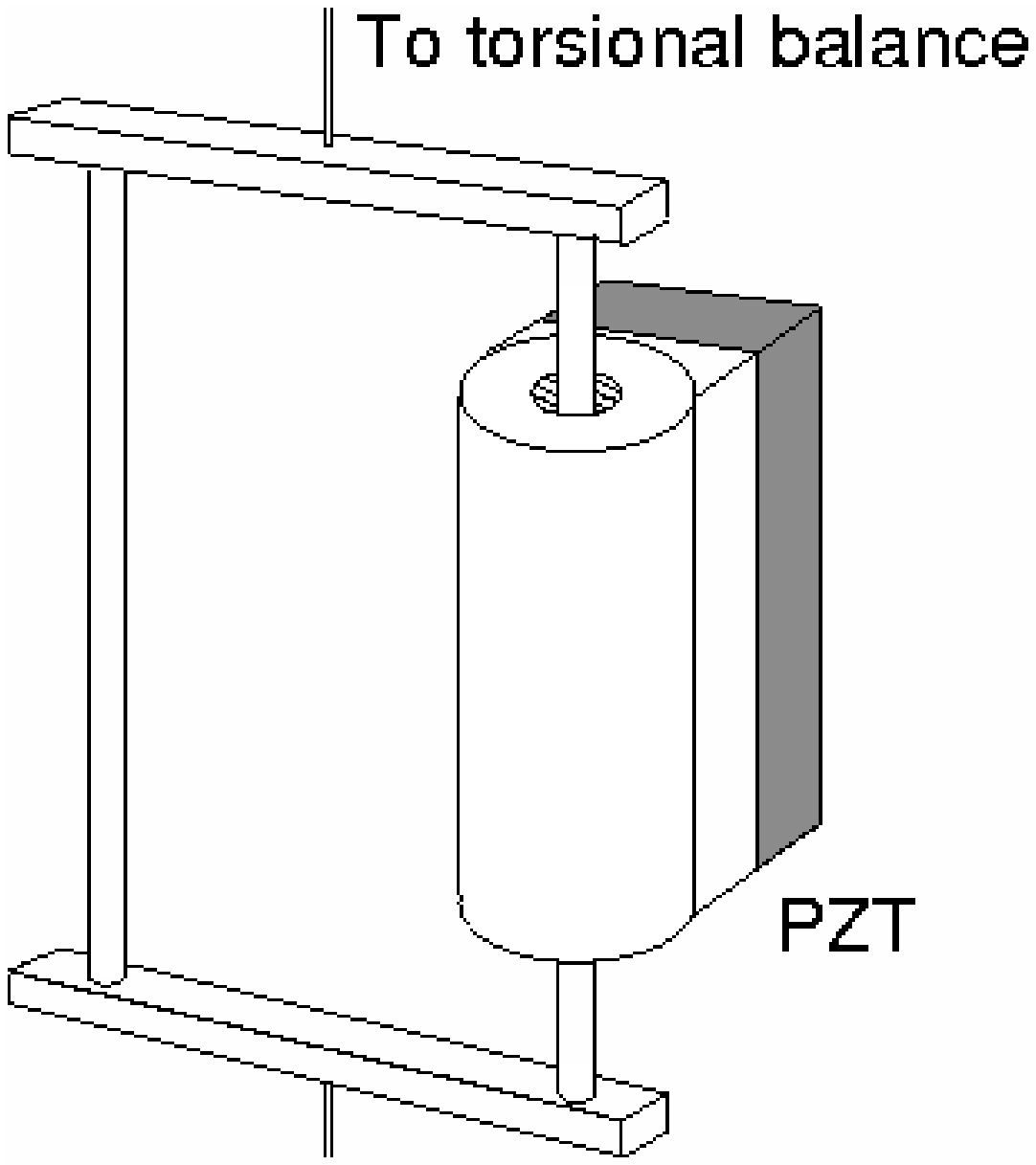,height=6cm,width=7cm,angle=0}}
\caption{Experimental setup for detecting Casimir forces using
quasi-concentric cylinders. The inner cylinder is rigidly
connected to a torsional balance and the signal to restore the
concentric configuration is measured after a controlled
displacement.} \label{fig1}
\end{figure}

We discuss possible experimental arrangements for measuring the
Casimir force between cylinders. As it was reported in the last years,
cylindrical shells provide a new and promising arena to study
Casimir interactions \cite{epl,pra05}.

\subsection{A null experiment}

Let us first consider the case of two eccentric cylindrical
shells, in an almost coaxial configuration. The concentric is an
unstable equilibrium position, so  one possibility is to repeat a
microscopic version of the experiment described in \cite{hoskins}
to test universal gravitation in the cm range, with a small
torsional balance mounted on the ends of the internal cylinder. In
this case the unstable force could be evidenced by intentionally
creating a controlled eccentricity and measuring the feedback
force required to bring the internal cylinder to zero
eccentricity, as depicted in Fig. 1 \cite{epl}.

This configuration has some advantages over the parallel plates
geometry. If there is no residual charge in the inner cylinder,
the system remains neutral and screened by the external one from
background noise sources, and from residual charges in the outer
cylinder. When the inner cylinder has a residual charge, there
will be a small potential difference between the cylinders, and
the coaxial configuration will be electrostatically unstable. The
electrostatic instability can be avoided by putting the cylinders
in contact, something ineluctable during the preliminary stages of
parallelization. Then the residual charge of the inner cylinder
will flow to the hollow cylinder, apart from a residual charge due
to the imperfections and finite length of the cylinders. This
residual charge will be smaller than for other geometries, as the
same discharging procedure does not work in the other
configurations (the efficiency of this procedure could be affected by the fact that
repeated contact between the cylindrical shells may result in a
degradation of the surfaces, as for instance an increased rugosity).

The electrostatic instability could also be
exploited to improve the parallelism between cylinders. One could
apply a time-dependent potential between the cylinders and measure
the force, as in the experiments to test the inverse-square
Coulomb law. Parallelism and concentricity would be maximum for a
minimum value of the force. Moreover, the expected gravitational
force is obviously null, this being an advantage if one looks for
intrinsically short-range extra-gravitational forces
\cite{onofrio06}.

\begin{figure}[t] 
\centerline{\includegraphics[width=11cm]{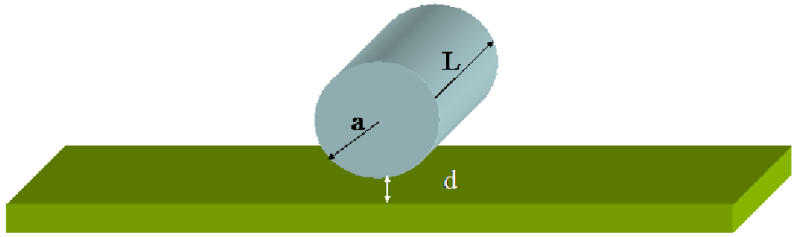}} 
\caption{Cylindrical shell in front of a conducting plane. Using PFA one can show that
the scaling of the Casimir force with distance is $\propto d^{-7/2}$ \cite{epl},
which is intermediate between
the plane-spherical ($\propto d^{-3}$) and the parallel plate 
configuration ($\propto d^{-4}$). This configuration is also intermediate for the 
absolute value of the force signal for typical values of the geometrical parameters.} 
\label{fig2}
\end{figure}

\subsection{A cylinder in front of a conducting plane}

Another possibility, which in principle is much more appealing
from the experimental point of view, is to consider the
configuration of a cylinder in front of a plane. The
cylinder-plane configuration of Fig. 2 is a compromise between the
different drawbacks and advantages of the parallel-plates and the
sphere-plate configurations. While this geometry offers a simpler
way to control the parallelism with respect to the parallel plate
case, at the same time gives rise to considerable force signal
which requires the study of the force at larger distances. Not
only can the study of the cylindrical-plane configuration provide
insights into the thermal contribution to the Casimir force
arising at any finite temperature, but also into the validity of
the PFA. This is because, as the signal
is stronger than in the sphere plane configuration, one can
envisage sufficiently precise measurements at larger distances.
This experiment is in progress \cite{pra05,jpa06}.

\subsection{Cylindrical rack and pinion}

Another interesting configuration is a cylindrical version of the
Casimir ``rack and pinion" proposed in Ref.\cite{golest}, in which
the rack is a corrugated cylindrical shell that encloses the
pinion, instead of a corrugated plane (see Fig. 3). In this case,
the cylindrical geometry is of interest for the opposite reason
than before: while reaching parallelism may be more complicated
than from the plane rack, the torque will be enhanced by a
geometric factor, and therefore could be more useful to generate
the motion of the pinion. Indeed, the interaction energy per unit
area between sinusoidally corrugated plane surfaces is given by
\cite{golest}

\begin{equation}
E_{pp}=\frac{\hbar c h^2}{d^5}\cos\left(\frac{2\pi x}{\lambda}\right)
J(\frac{d}{\lambda}),
\end{equation}
where $d$ is the mean distance between the plates, $h$ is the amplitude of the
corrugations and $x$ is the lateral displacement. We will not need the
explicit form of the function $J$.

\begin{figure}
\centerline{\psfig{figure=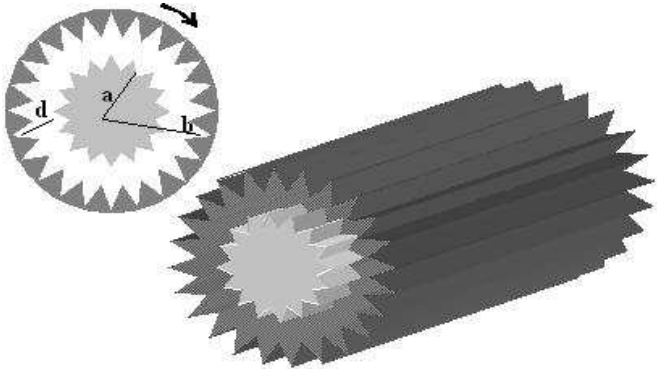,height=5cm,width=9cm,angle=0}}
\caption{Cylindrical version of the Casimir "rack and pinion". Rotation
of the outer cylinder induces a torque on the inner corrugated cylinder
due to the lateral Casimir force.}
\label{fig3}
\end{figure}

The interaction energy for the plane and for the cylindrical rack and
pinion can be easily
computed using the PFA. In the first case it is given by
\begin{equation}
E_{prp}=\hbar c h^2\cos\left(\frac{2\pi x}{\lambda}\right)L
a \int d\theta \frac{J(\frac{d(\theta)}{\lambda})}{d^5(\theta)},
\label{prp}
\end{equation}
where $a$ and $L$ are the radius and length of the cylinder, respectively,
and $d(\theta)=d+a(1-\cos\theta)$.

For the cylindrical case we have, instead, $E_{crp}=2\pi a
L E_{pp}$. This simple result is valid when the radii of the
outer ($b$) and the inner ($a$) cylinders satisfy $a\simeq b \gg d$.
Note that this configuration maximizes the
superposition between the two  surfaces, and therefore a uniform
rotation of the external shell will produce a torque on the pinion much larger
than the one produced by a plane rack moving with uniform velocity.

The ratio of the forces can be easily estimated from the ratio of
the interaction energies. Assuming that
$J(\frac{d(\theta)}{\lambda})$ is a smooth function, and that the
integral in Eq. (\ref{prp}) is dominated by $\theta\approx 0$, we
obtain $F_{crp}/F_{prp} \gtrsim \sqrt{a/d}\gg 1$.

%%%%%%%%%%%%%%%%%%%%%%%%%%%%%%%%%%%%%%%%%

\section{The exact formula for eccentric cylinders}

The evaluation of the Casimir interaction energy between eccentric
cylindrical shells (Fig. 4) has been initially performed using the
PFA in Ref.\cite{epl}, where the force between a cylinder and a
plate has also been computed and discussed for the first time
using the same approximation. However, it is possible to go beyond
the PFA, and find an exact formula for the interaction energy
\cite{pra06,njp06}. This can be done using a mode by mode
summation technique combined with the argument theorem.

\begin{figure}
\centerline{\psfig{figure=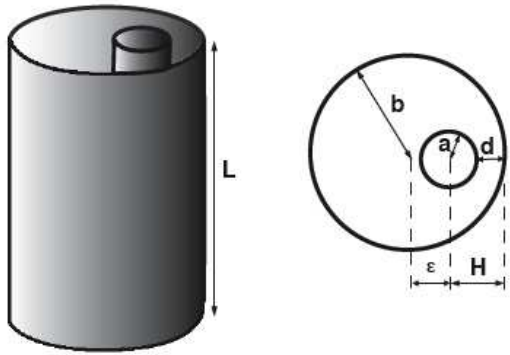,height=5cm,width=7cm,angle=0}}
\caption{Geometrical configuration for the eccentric cylinders. Two perfectly
conducting cylinders of radii $a < b$, length $L$, and eccentricity $\epsilon$
interacts via the Casimir force.}
\label{fig4}
\end{figure}

We start by expressing the Casimir energy as $E=(\hbar/2) \sum_p
(\omega_p - \tilde{\omega}_p)$, where $\omega_p$ are the
eigenfrequencies of the electromagnetic field satisfying perfect
conductor boundary conditions on the cylindrical surfaces, and
$\tilde{\omega}_p$ are the corresponding ones to the reference
vacuum (cylinders at infinite separation). In cylindrical
coordinates, the eigenmodes are $h_{n,k_z} = R_n(r,\theta) \exp[
-i( \omega_{n,k_z} t - k_z z)]$, where $\omega_p = \omega_{n,k_z}
= \sqrt{k_z^2 + \lambda_n^2}$, and $R_n$ ($\lambda_n$) are the
eigenfunctions (eigenvalues) of the 2D Helmholtz equation. Using
the argument theorem the sum over eigenmodes can be written as an
integral over the complex plane, with an exponential cutoff for
regularization. In order to determine the part of the energy that
depends on the separation between the two cylinders it is
convenient to subtract the self-energies of the two isolated
cylinders, $E_{12} (a,b,\epsilon )= E - E_1(a) - E_1(b)$. Then the
divergencies in $E$ are cancelled out by those ones in $E_1(a)$
and $E_1(b)$, and the final result for the interaction energy is
\begin{equation}
E_{12}(a,b,\epsilon ) = \frac{\hbar c L}{4 \pi} \int_0^{\infty} dy
\; y \; \log M(i y). \label{exactenergy}
\end{equation}
Here $M = (F/F_{\infty}) / [(F_1(\infty) / F_1(a)) \;
(F_1(\infty)/F_1(b)) ]$. The function $F$ is analytic and vanishes
at all the eigenvalues $\lambda_n$ ($F_{\infty}$, at
$\tilde{\lambda}_n$), and, similarly, $F_1$ vanishes for all
eigenvalues of the isolated cylinders. The function $M$ is the
ratio between a function corresponding to the actual geometrical
configuration and the one with the conducting cylinders far away
from each other. It is convenient to subtract a configuration of
two cylinders with very large and very different radii, while
keeping the same eccentricity of the original configuration. Eq.
(\ref{exactenergy}) is valid for two perfect conductors of any
shape, as long as there is translational invariance along the $z$
axis.

The solution of the Helmholtz equation in the annulus region
between eccentric cylinders has been considered in the framework
of classical electrodynamics and fluid dynamics. The
eigenfrequencies for Dirichlet boundary conditions (TM modes) and
for Neumann boundary conditions (TE modes) are given by the zeros
of the determinants of the non-diagonal matrices
\begin{eqnarray}
Q^{\rm TM}_{mn}&=& \left[J_n(\lambda a) N_m(\lambda b) -
J_m(\lambda b) N_n(\lambda a) \right]
J_{n-m}(\lambda \epsilon) ,  \nonumber \\
Q^{\rm TE}_{mn}&=& \left[J'_n(\lambda a)N'_m(\lambda b) -
J'_m(\lambda b) N'_n(\lambda a)\right] J_{n-m}(\lambda \epsilon) ,
\nonumber
\end{eqnarray}
where $J_n$ and $N_n$ are Bessel functions of the first kind. The
function $M$ can be written as $M = M^{\rm TE} M^{\rm TM}$, where
$M^{\rm TM}$ is built with
\begin{eqnarray}
&&F^{\rm TM} = {\rm det}
\left[ Q^{\rm TM}(a,b,\epsilon)Q^{\rm TM}(b,R,0) \right] \prod_n J_n(\lambda a), \nonumber \\
&&F_1^{\rm TM}(a) = {\rm det} \left[ Q^{\rm TM}(a,R,0) \right]
\prod_n J_n(\lambda a),
\end{eqnarray}
and $R$ is a very large radius. Similar expressions hold for
$M^{\rm TE}$.

The Casimir energy can be decomposed as a sum of TE and TM
contributions
\begin{equation}
E_{12} = \frac{\hbar c L}{4 \pi a^2} \int_0^{\infty} d\beta \beta
\left[ \log M^{\rm TE} \left( \frac{i\beta}{a} \right) + \log
M^{\rm TM} \left( \frac{i\beta}{a} \right) \right] \label{exact}
\end{equation}
with $M^{\rm TE,TM}(\frac{i\beta}{a}) = {\rm det} [ \delta_{np} -
A^{\rm TE,TM}_{np}]$. The  non-diagonal matrices $A^{\rm TE}_{np}$
and $A^{\rm TM}_{np}$ are
\begin{eqnarray}
A_{np}^{\rm TM}  &=& \frac{I_n(\beta)}{K_n(\beta)} \sum_m \frac{K_m(\alpha\beta)}{I_m(\alpha\beta)}
I_{m-n}(\beta \delta) I_{m-p}(\beta \delta) , \nonumber \\
A_{np}^{\rm TE}  &=& \frac{I'_n(\beta)}{K'_n(\beta)} \sum_m
\frac{K'_m(\alpha \beta)}{I'_m(\alpha \beta)} I_{m-n}(\beta
\delta) I_{m-p}(\beta \delta) \nonumber .
\end{eqnarray}
Here $I_n$ and $K_n$ are modified Bessel functions of the first
kind, $\alpha = b/a$ and $\delta = \epsilon /a$. The determinants
are taken with respect to the integer indices $n,p = -\infty,
\ldots, \infty$, and the integer index $m$ runs from $-\infty$ to
$\infty$. Eq.(\ref{exact}) is the exact formula for the
interaction Casimir energy between eccentric cylinders.

\begin{figure}
\centerline{\psfig{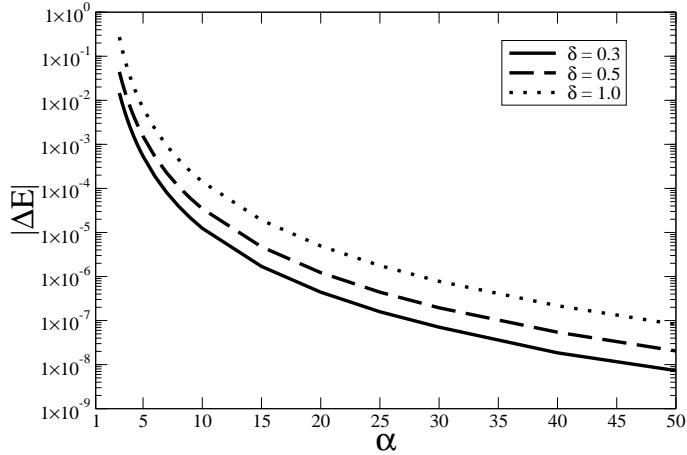}}
\caption{Exact Casimir interaction energy difference $\vert \Delta E\vert$ between
the eccentric and concentric configurations as a function of $\alpha = b/a$ for
different values of $\delta = \epsilon/a$. Energies are measured in units of
$L/4\pi a^2$.}
\label{fig5}
\end{figure}

This expression is rather complex to evaluate numerically, since
each term in the infinite matrix is a series involving Bessel
functions. However, we have been able to numerically evaluate the
exact Casimir interaction energy between eccentric cylinders as a
function of $\alpha$ for different values of the eccentricity
\cite{njp06}. We have calculated this energy as a function of
$\alpha$ for values that interpolate between the PFA (small
$\alpha$) values, and the asymptotic behavior for large $\alpha$
(see Fig. 5). The numerical convergence is better as $\alpha$
increases, while bigger matrices and more terms in each matrix
element are needed as $\alpha$ gets closer to $1$.

An interesting property of the exact formula is that 
reproduces the exact Casimir interaction energy
for the cylinder-plane configuration as a limiting case. 
Indeed, the eccentric
cylinder configuration tends to the cylinder-plane configuration
for large values of both the eccentricity $\epsilon$ and the
radius $b$ of the outer cylinder, keeping the radius $a$ of the
inner cylinder and the distance $d$ between the cylinders fixed.
Using the addition theorem  and uniform expansions for Bessel
functions it can be proved that, for $x \gg h$,
\begin{eqnarray}
\sum_m \frac{K_m(x+h)}{I_m(x+h)} I_{n-m}(x) I_{p-m}(x) \approx K_{n+p}(2h) , \nonumber \\
\sum_m \frac{K'_m(x+h)}{I'_m(x+h)} I_{n-m}(x) I_{p-m}(x) \approx -
K_{n+p}(2h) . \nonumber
\end{eqnarray}
Using  these equations (with $x\equiv \beta \epsilon/a$ and
$h\equiv \beta H/a$) in our exact formula we find
\begin{eqnarray}
A_{np}^{\rm TM, c-p} &=& \frac{I_n(\beta)}{K_n(\beta)} \; K_{n+p}(2 \beta H/a) , \nonumber \\
A_{np}^{\rm TE, c-p} &=& - \frac{I'_n(\beta)}{K'_n(\beta)} \;
K_{n+p}(2 \beta H/a) , \nonumber
\end{eqnarray}
which is the known result for the Casimir energy in the
cylinder-plane configuration \cite{Emig2006}.

%%%%%%%%%%%%%%%%%%%%%%%%%%%%%%%%%%%%%%%%%

\section{Concentric cylinders: new analytic and numerical results}

The exact formula for eccentric cylinders coincides, of course,
with the known result for the Casimir energy for concentric
cylinders ($\epsilon = 0$). Indeed, as $I_{n-m}(0) = \delta_{nm}$, in this
particular case the matrices $A_{np}^{{\rm TE,TM}}$ become
diagonal and the exact formula reduces to \cite{Mazzitelli2003}:
\begin{equation}
E_{12}^{\rm cc} = {L \over 4\pi a^2} \int_{0}^{\infty} d\beta \
\beta\ln M^{\rm cc}(\beta), \label{cc}
\end{equation}
where
\begin{equation}
M^{\rm cc}(\beta)=\prod_n \left[1-{I_n(\beta)K_n(\alpha
\beta)\over I_n(\alpha \beta)K_n(\beta)}\right]
\left[1-{I'_n(\beta)K'_n(\alpha \beta) \over I'_n(\alpha
\beta)K'_n(\beta)}\right] . \label{Mcc}
\end{equation}
The first factor corresponds to Dirichlet (TM) modes and the
second one to Neumann (TE) modes. The concentric-cylinders
configuration is interesting from a theoretical point of view,
since it can be used to test analytic and numerical methods. It
also has potential implications for the physics of nanotubes
\cite{pra06,klim}.

The short distance limit $\alpha - 1\ll 1$ has already been analyzed
for this case \cite{Mazzitelli2003}, and involves the summation over all values of $n$.
As expected, the resulting value is equal to the one obtained via
the proximity approximation, namely
\begin{equation}
E_{12, {\rm PFA}}^{\rm cc} = - \frac{\pi^3 L}{360 a^2} \;
\frac{1}{(\alpha-1)^3}. \label{proxi}
\end{equation}

In the opposite limit ($\alpha \gg 1$) it is easy
to prove that to leading order only the $n=0$ term contributes to
the interaction energy, and the energy decreases logarithmically
with the ratio $\alpha = b/a$,
\begin{equation}
E_{12}^{\rm cc} \approx - {1.26 L \over 8\pi b^2\ln\alpha}.
\label{cclargea}
\end{equation}

It is worth noticing that, while for small values of $\alpha$ both
TM and TE modes contribute with the same weight to the interaction
energy, the TM modes dominate in the large $\alpha$ limit.

\subsection{Beyond proximity approximation: the next to next to leading order}

In this Section we will compute analytic corrections to the PFA
given in Eq.(\ref{proxi}). Due to the simplicity of this configuration, 
we will be able to obtain not only the next to
leading order, but also the next to next to leading contribution.
In order to do that, we need the uniform expansion of the Bessel
functions. For example, we write

\begin{equation}
\frac{K_n(n \alpha y)}{K_n(n y)} = \frac{\sqrt{1 + y^2}}{\sqrt{1 +
\alpha^2 y^2}}\frac{(1 - \frac{u(t_\alpha )}{n})}{(1 -
\frac{u(t_1)}{n})} e^{n\left[\eta (\alpha y) - \eta (y)\right]},
\end{equation}
where
\begin{equation}
\eta (y)= \sqrt{1 + y^2}+ \ln{\frac{y}{1 + \sqrt{1 + y^2}}} ~;~
u(t) = \frac{3 t - 5 t^3}{24}~;~ t_\alpha = \frac{1}{\sqrt{1 +
\alpha^2 y^2}},
\end{equation}
and similar expressions for the functions $I_n$.

With these expansions at hand, we can evaluate the the matrix $M$
both for the TE and TM modes. After a long calculation, 
it is possible to show that the Casimir energy, beyond the
proximity approximation can be written as

\begin{equation}
E_{12}^{\rm cc} \approx -\frac{\pi^3 L}{360 a^2(\alpha -
1)^3}\left\{1 + \frac{1}{2}(\alpha - 1) - \left(\frac{2}{\pi^2} +
\frac{1}{10} \right)(\alpha - 1)^2 + ...\right\}
.\label{ntntlead}\end{equation}

In the expression above, the first term inside the parenthesis
corresponds  to the proximity approximation contribution in
Eq.(\ref{proxi}), while the second and third terms are the first
and second order corrections, respectively. It is important to
stress here that both TM and TE modes contribute with the same
weight to the energy up to the next to leading order, but it is
not the case in the third term. There is a factor  $1/\pi^2$
coming from the TM mode, and a factor $1/\pi^2 + 1/10$
corresponding to the TE one.

\subsection{Improving the convergence of the numerical evaluation}

Numerical calculations of the Casimir energy for $\alpha$ very
close to one are really difficult since  big number of terms
have to be considered in the sums, and therefore convergence problems
arise.

In order to perform a numerical evaluation of the Casimir energy
for the concentric cylinder case, we will describe a subtraction
method, in which we have used the proximity approximation value of
the energy to improve numerics.

As a guiding example, let us consider the following sum:
\begin{equation}
 z_M=\sum_{n=1}^M \frac{1}{n^{1.1}}.
\end{equation}
The convergence of this sum as $M\rightarrow\infty$ is extremely
slow, as the following numbers suggest: $z_{10^3}=5.5728$,
$z_{10^5}=7.4222$ and $z_{\infty}=10.5844$. About $10^{20}$ terms
are needed to get an accuracy of $1\%$.

Then, we add and subtract the function $\int_1^{M} \frac{dx}{x^{1.1}}$, that 
reproduces the behaviour of the series when the exponent in the denominator is 
close to one. In this way,
\begin{equation}
 z_M= z_M -\int_1^{M} \frac{dx}{x^{1.1}}+\int_1^{M} \frac{dx}{x^{1.1}}=
D_M + \int_1^{M} \frac{dx}{x^{1.1}}\rightarrow D_\infty + 10.
\end{equation}
The convergence of the new function $D_M$ is notably faster as
$D_{10}=0.6234$ and $D_{1000}=0.5847$, i.e. $1\%$ accuracy is
obtained with less than $10$ terms.

\begin{figure}
\centerline{\psfig{figure=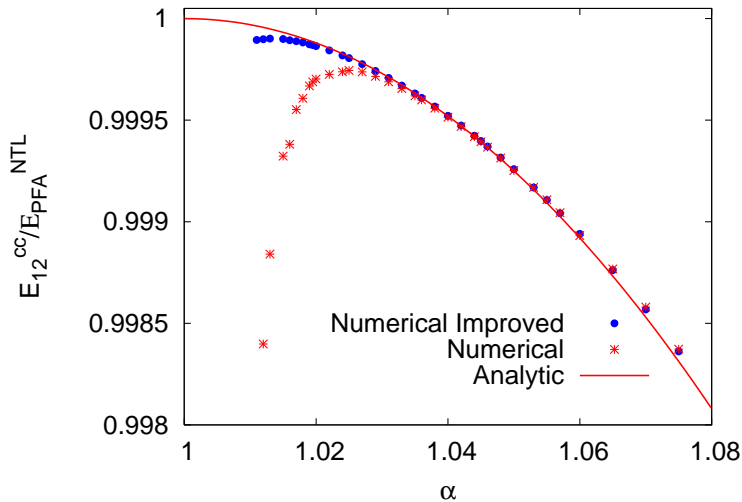,height=7cm,width=10cm,angle=0}}
\caption{Ratio between the exact Casimir energy for concentric cylinders
 $E_{12}^{cc}$ and the Casimir energy estimated using the PFA up to the next 
to leading order $E_{PFA}^{NTL}$, as a function of
the parameter $\alpha$. This is done
for two different methods: the numerical (of slow convergence) 
and the numerical improved (subtraction method). Solid line 
indicates the analytic result from Eq.(\ref{ntntlead}).}
\label{fig6}
\end{figure}

In the case we are concerned here, we can add and subtract the
interaction energy for concentric cylinders computed using the
leading uniform asymptotic expansion for the Bessel functions, up
to first order in $\alpha-1$:
\begin{equation}
\frac{K_n(n \alpha y)}{K_n(n y)} \frac{I_n(n y)}{I_n(n \alpha y)} \simeq e^{-2 n(\alpha - 1)\sqrt{1+y^2}}.
\end{equation}
We denote by $\tilde{E}$ the interaction
energy obtained by inserting these expansions into Eq. (\ref{cc}). Now we write
\begin{equation}
 E_{12}^{cc}=(E_{12}^{cc} - \tilde{E}) +\tilde{E}. \label{Emodificada}
\end{equation}
The last term in the above expression is easily written into an
analytic expression, which contains the leading order of the
Casimir energy. Meanwhile, the difference contained in the
brackets in Eq. (\ref{Emodificada}), has a faster convergence than
the original sum and therefore, can be easily calculated
numerically.

In this context, in Fig.\ref{fig6} we present both Casimir
energy of the concentric cylinders for the direct numerical calculation (of
slow convergence) and the alternative method mentioned above. In this figure 
we plot the ratio $E_{12}^{cc}/E_{PFA}^{NTL}$ where 

\begin{equation}
E_{PFA}^{NTL} = -\frac{\pi^3 L}{360 a^2(\alpha -
1)^3}\left\{1 + \frac{1}{2}(\alpha - 1)\right\}.
\end{equation}
As can be seen, with this subtraction method it is possible to
compute the exact energy for values of $\alpha$ much closer to
$1$, while the accuracy of the direct calculation is worse for 
$\alpha < 1.02$~.

A similar method could in principle be applied to the eccentric
cylinders or the cylinder-plane configurations, although in
these cases the main difficulty is the analytic evaluation of the
approximate energy to be added and subtracted.

%%%%%%%%%%%%%%%%%%%%%%%%%%%%%%%%%%%%%%%%%
\section{Final remarks}

In this paper we have described experimental proposals and
theoretical aspects of the Casimir interaction energy between
cylindrical shells. We have reviewed previous works on the
subject, in which we obtained an exact formula for the interaction
energy between eccentric cylinders \cite{pra06,njp06}, and where we 
discussed the advantages of considering experiments involving cylinders
\cite{epl,pra05}.

We have also presented some new results. In particular, we have
shown that a cylindrical version of the non contact rack and pinion
powered by the lateral Casimir force, proposed in Ref.\cite{golest}, 
would have a larger torque because of a geometric enhancement.

From a theoretical point of view, we have found an analytic
formula for the interaction energy between concentric cylinders
beyond the PFA, including first and second order corrections. We
have also presented a subtraction method useful to improve the
convergence of the numerical calculations as the concentric
surfaces get closer to each other. We hope to generalize these
results to other geometries in future works.

%%%%%%%%%%%%%%%%%%%%%%%%%%%%%%%%%%%%%%%%%

\ack This work has been supported by CONICET, UBA and ANPCyT,
Argentina.

%%%%%%%%%%%%%%%%%%%%%%%%%%%%%%%%%%%%%%%%%

\end{document}